\documentclass[a4paper,fleqn,usenatbib]{mnras}

\usepackage[T1]{fontenc}
\usepackage{ae,aecompl}

\usepackage{graphicx}	
\usepackage{amsmath}	
\usepackage{amssymb}	

\title[A deep Seyfert-2 galaxy at $z=0.222$ behind NGC\,300]{Discovery of a deep Seyfert-2 galaxy at \boldmath${z=0.222}$ behind NGC\,300}

\author[J. A. Combi et al.]{
J. A. Combi,$^{1,2}$\thanks{E-mail: jcombi@fcaglp.unlp.edu.ar}
F. Garc\'{\i}a,$^{1,2}$
M. J. Rodr\'iguez$^{3}$
R. Gamen$^{2,3}$
and S. A. Cellone$^{2,3}$
\\
$^{1}$Instituto Argentino de Radioastronom\'{\i}a (CCT La Plata, CONICET), C.C.5, (1894) Villa Elisa, Buenos Aires, Argentina.\\
$^{2}$Facultad de Ciencias Astron\'omicas y Geof\'{\i}sicas, Universidad Nacional de La Plata, Paseo del Bosque, B1900FWA La Plata, Argentina.\\
$^{3}$Instituto de Astrof\'isica de La Plata, (CCT La Plata, CONICET, UNLP), Paseo del Bosque, B1900FWA, La Plata, Argentina.}

\date{Accepted XXX. Received YYY; in original form ZZZ}

\pubyear{2016}

\begin{document}
\label{firstpage}
\pagerange{\pageref{firstpage}--\pageref{lastpage}}
\maketitle

\begin{abstract}
We report  on the unveiling of the nature of the unidentified X-ray source 3XMM\,J005450.3--373849 as a Seyfert\,2 galaxy located behind the spiral galaxy NGC\,300 using \emph{Hubble} Space Telescope data, new spectroscopic Gemini observations and available XMM-\emph{Newton} and \emph{Chandra} data. We show that the X-ray source is positionally coincident with an extended optical source, composed by a marginally resolved nucleus/bulge, surrounded by an elliptical disc-like feature and two symmetrical outer rings. The optical spectrum is typical of a Seyfert\,2 galaxy redshifted to $z=0.222 \pm 0.001$, which confirms that the source is not physically related to NGC\,300. At this redshift the source would be located at 909$\pm$4 Mpc (comoving distance in the standard model). The X-ray spectra of the source are well-fitted by an absorbed power-law model. By tying $N_\mathrm{H}$ between the six available spectra, we found a variable index $\Gamma$ running from $\sim$2 in 2000-2001 years, to 1.4--1.6 in the 2005-2014 period. Alternatively, by tying $\Gamma$, we found variable absorption columns of $N_\mathrm{H} \sim 0.34 \times 10^{-22}$~cm$^{-2}$ in 2000-2001 years, and $0.54-0.75 \times 10^{-22}$~cm$^{-2}$ in the 2005-2014 period. Although we cannot distinguish between an spectral or absorption origin, from the derived unabsorbed X-ray fluxes, we are able to assure the presence of long-term X-ray variability. Furthermore, the unabsorbed X-ray luminosities of $0.8-2 \times$10$^{43}$~erg~s$^{-1}$ derived in the X-ray band are in agreement with a weakly obscured Seyfert-2 AGN at $z\approx0.22$.
\end{abstract}




\section{Introduction}

Unidentified extragalactic variable X-ray sources with extended emission at radio and optical wavelengths are often associated to Active Galactic Nuclei \citep[AGNs][]{AGNxrayARAA} or Ultraluminous X-ray sources \citep[ULXs, e.g.][]{Fabb-2006}. Multiwavelength analyses, from radio to gamma-ray are needed to disentangle and characterise the different components and nature of such astrophysical systems. 

Of particular interest is the X-ray source 3XMM J005450.3$-$373849, located within the spiral galaxy NGC\,300, belonging to the nearby Sculptor Group of galaxies. It was first studied by \citet{RP-2001}, who suggested an X-ray binary nature for the source, based on hardness-ratio values and its variable behaviour obtained from observations performed with the High Resolution Imager of \emph{ROSAT}. Later on, \citet{PFP+2004} showed that the source (named XMM5 in their work) presents a radio counterpart (ATCA J005450.3$-$373850), whose extended radio emission may be the result of synchrotron radiation from jets, possibly associated to an extragalactic microquasar. A study of the global properties of X-ray point sources in NGC\,300, and their optical counterparts was performed by \citet{CWSK-2005}, who associated the source (named Source 7 in their work) with an apparently extended optical counterpart. 

In this paper, we report results of a combined optical and X-ray analysis using archival data of the HST, XMM-\emph{Newton} and \emph{Chandra} observatories, together with new optical spectroscopic \emph{Gemini} observations, of the extragalactic X-ray source 3XMM J005450.3$-$373849, and we provide convincing evidence about its true nature as a Seyfert-2 AGN galaxy located at $z=0.222$. The structure of the paper is as follows: in Sect.\,2 we describe \emph{Hubble}, \emph{Gemini}, XMM-\emph{Newton} and \emph{Chandra} observations and data reduction. Optical/X-ray analysis and results are shown in Sect.\,3. Finally, in Sect.\,4, we discuss the implications of our results and summarise our main conclusions.

\begin{table*}
 \begin{centering}
  \caption{\emph{Chandra} and XMM-\emph{Newton} observations of 3XMM J005450.3--373849.}
  \begin{tabular}{c l cccccc}
  \hline
Obs. &Satellite &    Observation ID &       Date             &Starting time & Exposure [ks] &GTI  [ks] & Observation Mode      \\
     &                                                   &                         &                     & &   MOS1/MOS2/pn                     &        & \\
 \hline
1 & XMM      & 0112800201      & 2000-12-26          &  18:14:32     &    33.7/33.7/28.3 & 32.3/31.9/24.5  & PFWE\\
2 & XMM      & 0112800101      & 2001-01-01          &   13:06:32    &    43.4/43.4/34.6   &  42.9/43.1/34.4  & PFWE\\
3 & XMM      & 0305860301      & 2005-11-25          &   07:20:47    &    36.0/36.0/30.8   & 35.6/35.9/30.5  & PFWE\\
4 & XMM       &  0656780401   & 2010-05-28          &   13:23:23    &   17.8/17.8/11.8 &  13.9/14.6/10.9   & PFWE\\
5 & Chandra  &  12238            &    2010-09-24        &     02:57:22  &   65.80 &  62.9  & VFAINT  \\
6 & Chandra  &  16028            &    2014-05-16        &    19:33:04   &    65.08 &  64.2    & VFAINT \\
\hline
\end{tabular}
\label{obstable}
\end{centering}
\\ 
\begin{flushleft}
All observations were taken from their respective mission archives. PFWE refers to the Prime Full Window Extended observation mode, and VFAINT to very faint mode.
\end{flushleft}
\end{table*}

\section{Observations and data reduction}

\subsection{X-ray data}

We analysed four XMM-\emph{Newton} and two \emph{Chandra} observations of 3XMM J005450.3$-$373849 obtained along a period of fourteen years between 2000/12/26 and 2014/05/16. The XMM-\emph{Newton} observations were performed with the European Photon Imaging Camera (EPIC), which consists of three detectors, two MOS cameras \citep{turner2001}, and one PN camera \citep{struder2001} operating in the 0.2$-$15~keV range. XMM-\emph{Newton} data were analysed with the XMM-\emph{Newton} Science Analysis System (SAS) version 14.0.0. Starting from level-1 event files, the latest calibrations were applied with the \textsc{emproc} and \textsc{epproc} tasks. For the MOS data, we selected only events with patterns 0 through 12 and applied flag filters XMMEA$_{-}$EM. For the PN data, we selected only events with patterns 0 through 4 and applied flag filters XMMEA$_{-}$EP. According to the light curves of MOS and PN cameras, we further excluded time-intervals with high background rates by setting good time interval (GTI) thresholds on 0.35~cts~s$^{-1}$ in the 0.3$-$12 keV range for MOS and 0.4~cts~s$^{-1}$ in the 0.3$-$15 keV range for PN, respectively.

\emph{Chandra} observations performed with the Advanced CCD Image Spectrometer (ACIS) camera were extracted from the archive. ACIS operates in the 0.1$-$10~keV range with high spatial resolution ($0.5''$). These observations were calibrated using the CIAO (version 4.7) and CALDB (version 4.6.7) packages by means of the {\sc chandra\_repro} task. Detailed information of the X-ray observations and instrumental characteristics are given in Table~\ref{obstable}.

\subsection{Optical data}

In order to analyze the properties of the optical counterpart of 3XMM J005450.3$-$373849, found by \citet{CWSK-2005}, we used optical images obtained with the Advanced Camera for Surveys (ACS) mounted in the HST. The observations correspond to the HST Cycle 11, and were obtained as part of the GO-9492 program (PI: F. Bresolin) from 2002 July to 2002 December. Three exposures of 360\,s are available, performed in the F435W, F555W and F814W filters \citep{BPGK-2005}, which are similar to the classical Johnson $BVI$ bands.  The ACS Wide Field Camera (WFC) has a mosaic of two CCDs detectors of $4096 \times 2048$ pixels and a scale of 0.049$''$/pixel, covering a field of $3.3' \times 3.3'$. These images and their corresponding photometric data \citep{dal2009}, were obtained from the STScI (MAST; \url{https://archive.stsci.edu/}) database. The photometry was carried out using the package DOLPHOT adapted for the ACS camera \citep{Dol2000}.

\begin{table}
\caption{Spectral parameters of 3XMM J005450.3--373849.}
\begin{centering}
{\bf Model I}
 
\begin{tabular}{cccc}
\hline
Obs. & $N_\mathrm{H}$ [$10^{22}$~cm$^{-2}$]     & $\Gamma$                       & Flux (0.5--8.0 keV)\\
\hline
1 &	 $0.40\pm0.03$  & $2.06\pm 0.08$ & $ 0.145\pm 0.006 $\\
2 &	 $\dagger$  	& $1.97\pm 0.08$ & $ 0.120\pm 0.004$\\
3 &	 $\dagger$  	& $1.65\pm 0.12$ & $0.057\pm 0.004$\\
4 &	 $\dagger$  	& $1.48\pm 0.13$ & $0.066\pm 0.004$\\
5 &	 $\dagger$  	& $1.51\pm 0.11$ & $ 0.111\pm 0.006 $\\
6 &	 $\dagger$  	& $1.49\pm 0.12$ & $ 0.102\pm 0.006 $\\
\hline
$\chi^2$ & 216.18 (226 d.o.f) &\\
\hline
\end{tabular}
{\bf Model II}

\begin{tabular}{cccc}
\hline
Obs. & $N_\mathrm{H}$ [$10^{22}$~cm$^{-2}$]      & $\Gamma$                       & Flux (0.5--8.0 keV)\\
\hline
1 &	 $0.33\pm0.03$  & $1.89\pm 0.06$ & $0.138\pm 0.006 $\\
2 &	 $0.36\pm0.03$ 	& $\dagger$ & $0.116\pm 0.005$\\
3 &	 $0.54\pm0.09$ 	& $\dagger$ & $0.062\pm 0.006$\\
4 &	 $0.66\pm0.11$	& $\dagger$ & $0.076\pm 0.007$\\
5 &	 $0.60\pm0.10$	& $\dagger$ & $ 0.119\pm 0.009 $\\
6 &	 $0.75\pm0.14$	& $\dagger$ & $ 0.121\pm 0.011 $\\
\hline
$\chi^2$ & 229.67 (226 d.o.f) &\\
\hline
\end{tabular}
\label{spectable}
\end{centering}
\\ Error values are 1-$\sigma$ (68\%) for every single parameter and unabsorbed fluxes in the 0.5-8.0 keV energy range are given in units of 10$^{-12}$~erg~cm$^{-2}$~s$^{-1}$.  
Parameters indicated with a $\dagger$ were tied between each other during the fit. Model I corresponds to values of $N_{\rm H}$ tied and Model II to values of $\Gamma$ tied, respectively.
\end{table}

\subsection{Gemini spectroscopy}

Spectra for 3XMM J005450.3$-$373849 were taken with the Gemini MultiObject Spectrograph (GMOS) at the Gemini South telescope under the program GS-2015B-DD-6 (PI: J. Combi).

We have adopted a single long-slit of width $0.75''$ to obtain the spectra of the target. We used the R400$_{-}$G5325 grism and obtained five exposures of 1200\,s in two different central wavelengths (three in 6700\,\AA\ and two in 6500\AA). This setup resulted in a spectral coverage of 4200--9050\,\AA\ and a resolution (defined as $\lambda/$FWHM measured in some emission lines of the CuAr arc lamps spectra) $R\sim1300$.

The flat-field frames were observed at the same position of the target. The CuAr arc lamps were observed as day-time calibrations, thus flexures can introduce some systematic uncertainties in the wavelength calibration which are not important to the aims of this work. The observations were reduced using the \textsc{gemini} package within \textsc{iraf}\footnote{\url{http://iraf.noao.edu/}}.

\section{Results}

\subsection{X-ray spectral and temporal analysis}

We checked the position of 3XMM J005450.3$-$373849 in Chandra observations running the \textsc{wavdetect} task which throwed $\alpha= 00^\mathrm{h}\,54^\mathrm{m}\,50\fs2$ and $\delta=-37\degr\,38\arcmin\,50\farcs6$ (J2000.0) with a positional uncertainty of $0.75''$. This value agrees very well with the values reported by \citet{RWW+2015} on their 3XMM-DR5 catalogue: $\alpha= 00^\mathrm{h}\,54^\mathrm{m}\,50\fs3$ and $\delta=-37\degr\,38\arcmin\,49\farcs5$ (J2000.0) with a $0.8''$ position error. We also used the \textsc{srcextent} CIAO task to calculate the size of the source, which resulted point-like at Chandra resolution.

In order to analyze the physical properties of the X-ray emission detected from 3XMM J005450.3$-$373849 in detail, we extracted spectra from circular regions centered at the position of the source for both XMM-Newton and Chandra observations. For XMM-\emph{Newton} data, spectra were obtained in the 0.5$-$12 keV energy range, for radii of $15''$ using \textsc{evselect} SAS task with the appropriate parameters for EPIC MOS~1/2 and PN cameras, and ancillary and response matrices were created using \textsc{arfgen} and \textsc{rmfgen} tasks, respectively. For the spectral extraction in MOS1/2 and PN cameras we selected events with FLAG == 0. In the case of \emph{Chandra} data, we extracted spectra using radii of $2''$ by means of the \textsc{specextract} CIAO task. For both telescopes, background spectra were extracted from annuli regions centered at the position of the source and the spectra were grouped with a minimum of 16 counts per bin. Spectral analysis was performed using the XSPEC package version 12.8 \citep{arnaud1996} in the 0.5--8.0~keV energy range, where the source is detected. 

We simultaneously fitted the spectra obtained with the XMM-\emph{Newton} and \emph{Chandra} detectors, using a power-law (PL) model modified by interstellar absorption \citep[\textsc{phabs};][]{balucinska1992}. In order to fit the data we used to alternatives. In Model I, we tied the hydrogen column density, $N_\mathrm{H}$, of all observations allowing to vary freely the spectral indices, $\Gamma$, and their normalizations, while, in Model II, we tied $\Gamma$, allowing to vary the $N_{\rm H}$. For Model I, the best fit model returned an hydrogen column density $N_\mathrm{H} =  0.40\pm0.03 \times 10^{22}$\,cm$^{-2}$ and power-law indices varying from $\sim$2, for Obs. 1 and 2, to 1.48--1.65 for Obs. 3 to 6, with a total $\chi^2$ of 216.18 for 226 d.o.f. Unabsorbed X-ray fluxes in the 0.5--8.0~keV energy range run from 0.145 to 0.057 $\times 10^{-12}$~erg~cm$^{-2}$~s$^{-1}$. For Model II, the best fit corresponds to a spectral index $\Gamma=1.89\pm0.06$, and absorption columns varying from $\sim 0.35 \times 10^{22}$~cm$^{-2}$, for Obs. 1 and 2, to $0.54-0.75 \times 10^{22}$\,cm$^{-2}$, for Obs. 3 to 6, with a total $chi^2$ of 229.67 for 226 d.o.f. Unabsorbed X-ray fluxes in the 0.5--8.0~keV range show a similar behaviour to Model I, being independent to the chosen model. The available data do not allow us to adequately fit the whole set of spectra leaving all the parameters untied, as there are no enough statistics. However, in both cases we observe the presence of a long-term X-ray variability. The full results of the fits are summarized in Table~\ref{spectable}.

\begin{figure*}
\centering
{\bf Model I \hspace{0.43\textwidth} Model II}

\vspace{-0.5cm}

\includegraphics[width=6cm,angle=270]{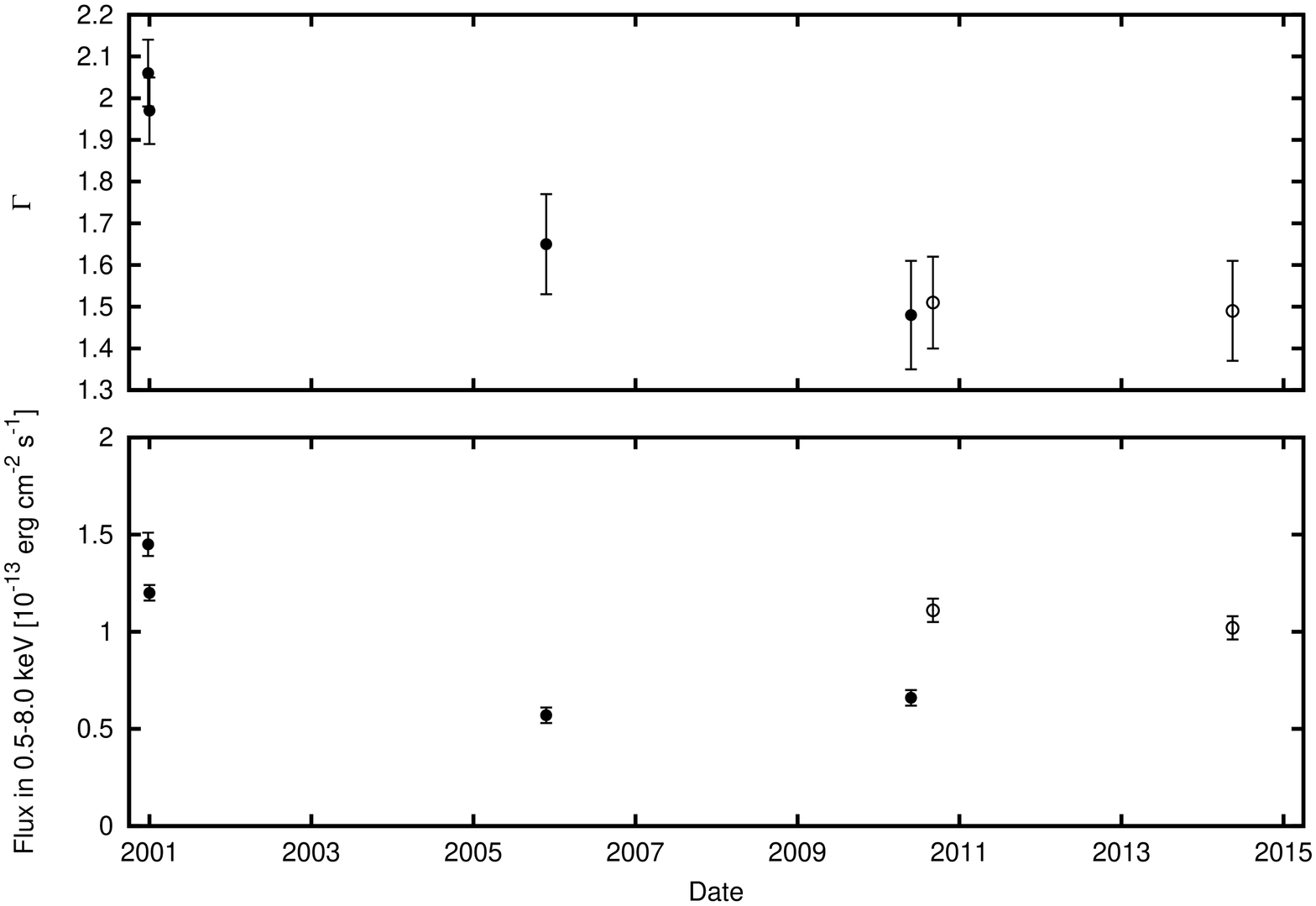}
\includegraphics[width=6cm,angle=270]{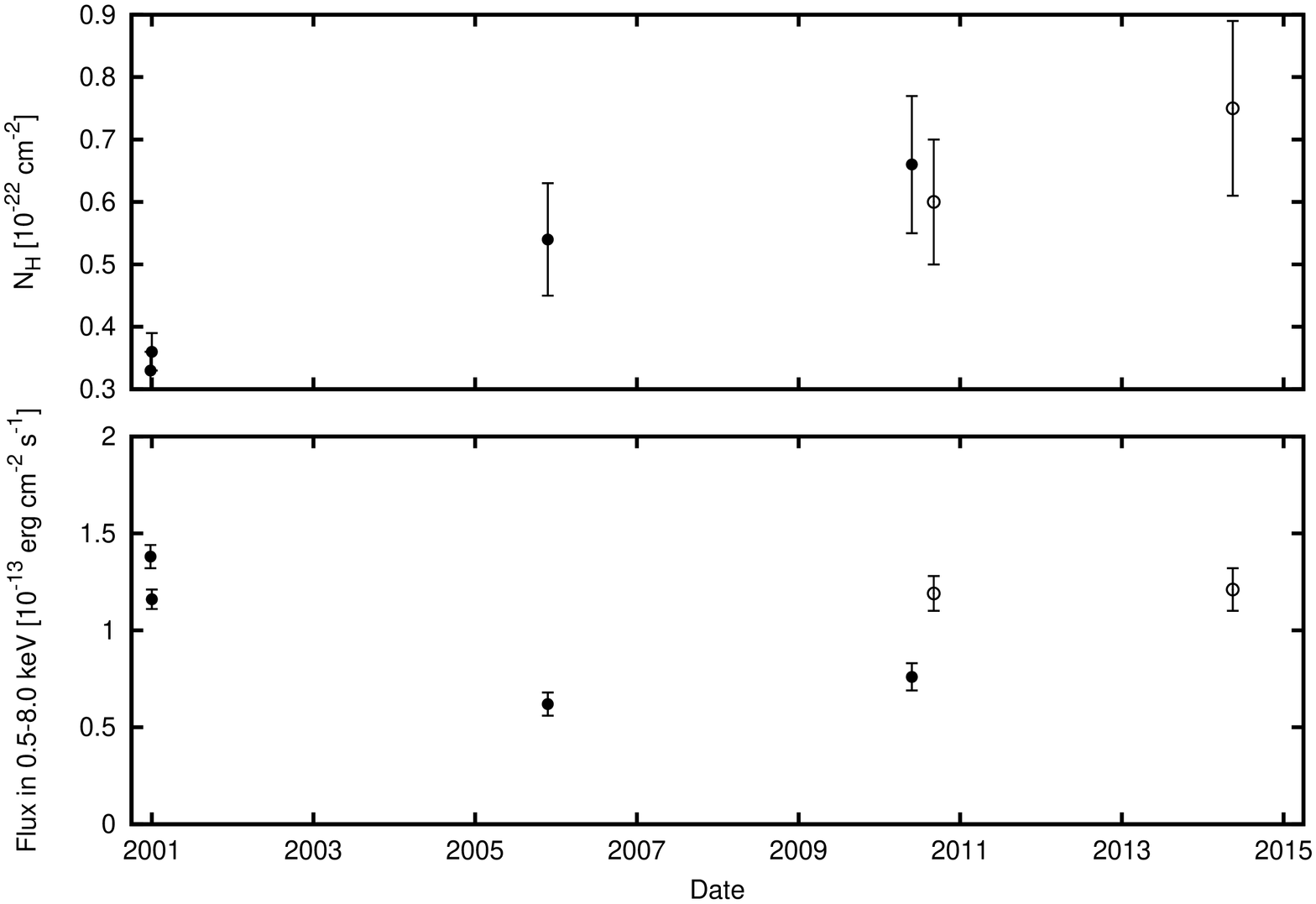}
\caption{\emph{Left panels:} Spectral results for Model I. \emph{Upper panel:} Spectral index $\Gamma$ evolution from 2000 Dec 26 to 2014 May 16. \emph{Bottom panel:} Unabsorbed total flux (0.5-8.0 keV) evolution in the same period. \emph{Right panels:} Spectral results for Model II. \emph{Upper panel:} $N_{\rm H}$ evolution from 2000 Dec 26 to 2014 May 16. \emph{Bottom panel:} Unabsorbed total flux (0.5-8.0 keV) evolution  also in the same period. Black and white circles represent XMM-\emph{Newton} and \emph{Chandra} observations, respectively.}
\label{fig_lc}
\end{figure*}

\begin{figure}
\centering
\includegraphics[height=0.45\textwidth,angle=270]{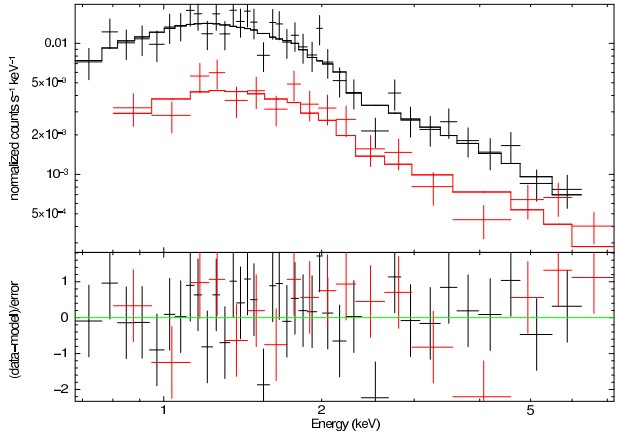}
\caption{XMM-\emph{Newton} PN spectra of 3XMM J005450.3$-$373849 obtained from observations 1 (black) and 4 (red). Solid lines represent the best fit using an absorbed powerlaw model (see Table~\ref{spectable}). Lower panel shows $\chi^2$ fit residuals. Error bars are at 90\% confidence intervals.}
\label{fig_spectra}%
\end{figure}

On the left panel of Figure~\ref{fig_lc} we plot the spectral results corresponding to Model I. On its upper panel, we present the evolution of the spectral index $\Gamma$ of 3XMM J005450.3$-$373849 from 2000 Dec 26 to 2014 May 16. The power-law spectral index was $\sim 2$ during the period 2000/2001, decreasing to $\sim 1.5$ in 2005/2014. On its lower panel we show the corresponding full 2000-2014 unabsorbed X-ray light curve in the 0.5-8.0~keV energy range. On the right panel of Figure~\ref{fig_lc} we display the spectral results from Model II. On its upper panel, we show the evolution of the absorption column $N_{\rm H}$, while on its lower panel we plot the corresponding X-ray light curve in the same ranges. Although a few years of data are missing (2001-2005), in both cases the source displays at least two different states. Regardless of the model chosen, the unabsorbed flux evolution shows the same behaviour which confirms the long-term X-ray variability of the source. To make this clear, on Figure~\ref{fig_spectra} we show background-subtracted PN spectra of XMM-\emph{Newton} observations 1 and 4 of 3XMM J005450.3$-$373849 as fitted by the absorbed power-law model.

\begin{figure*} %
\centering
\includegraphics[width=5.8cm,angle=0]{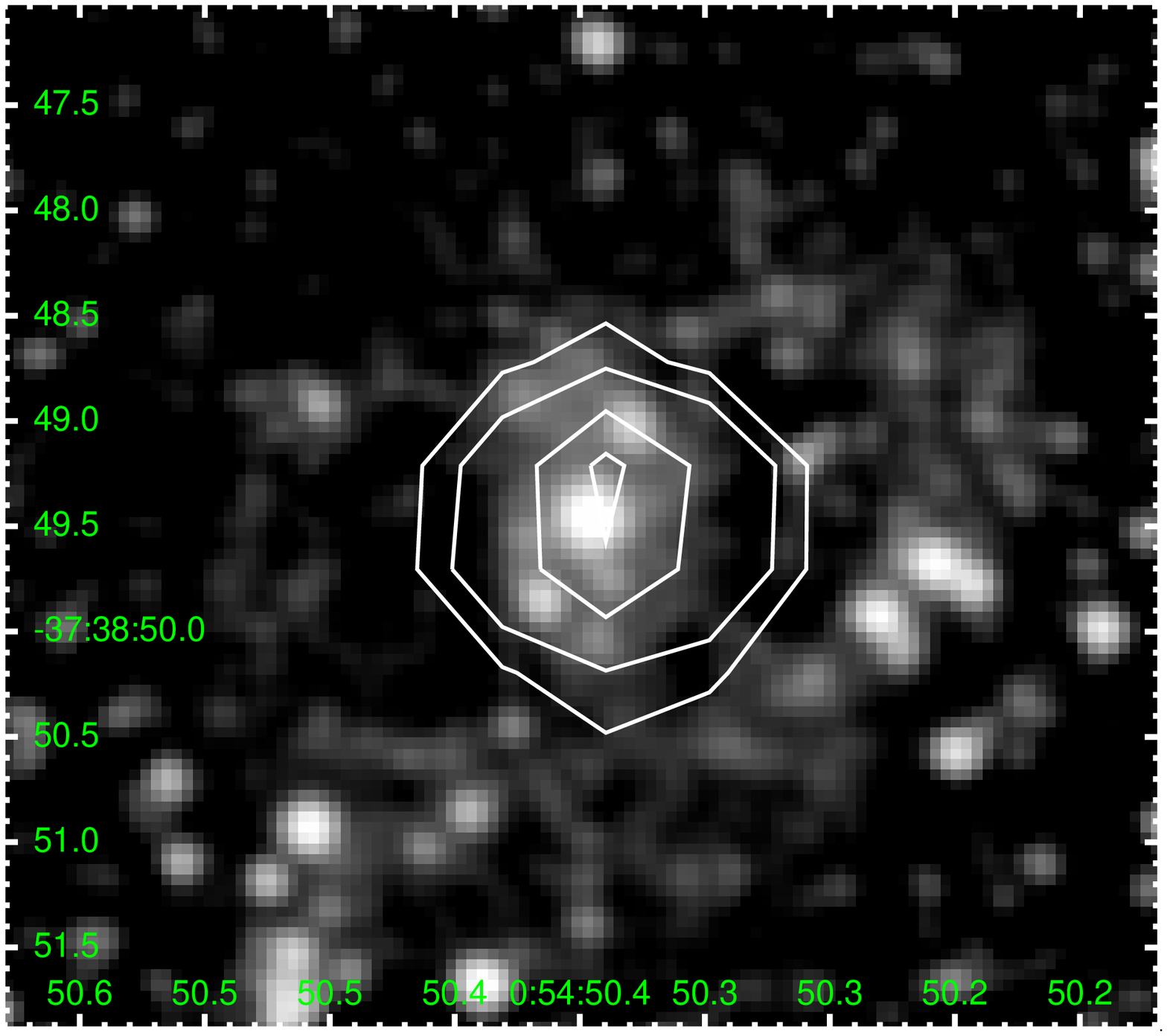}
\includegraphics[width=5.8cm,angle=0]{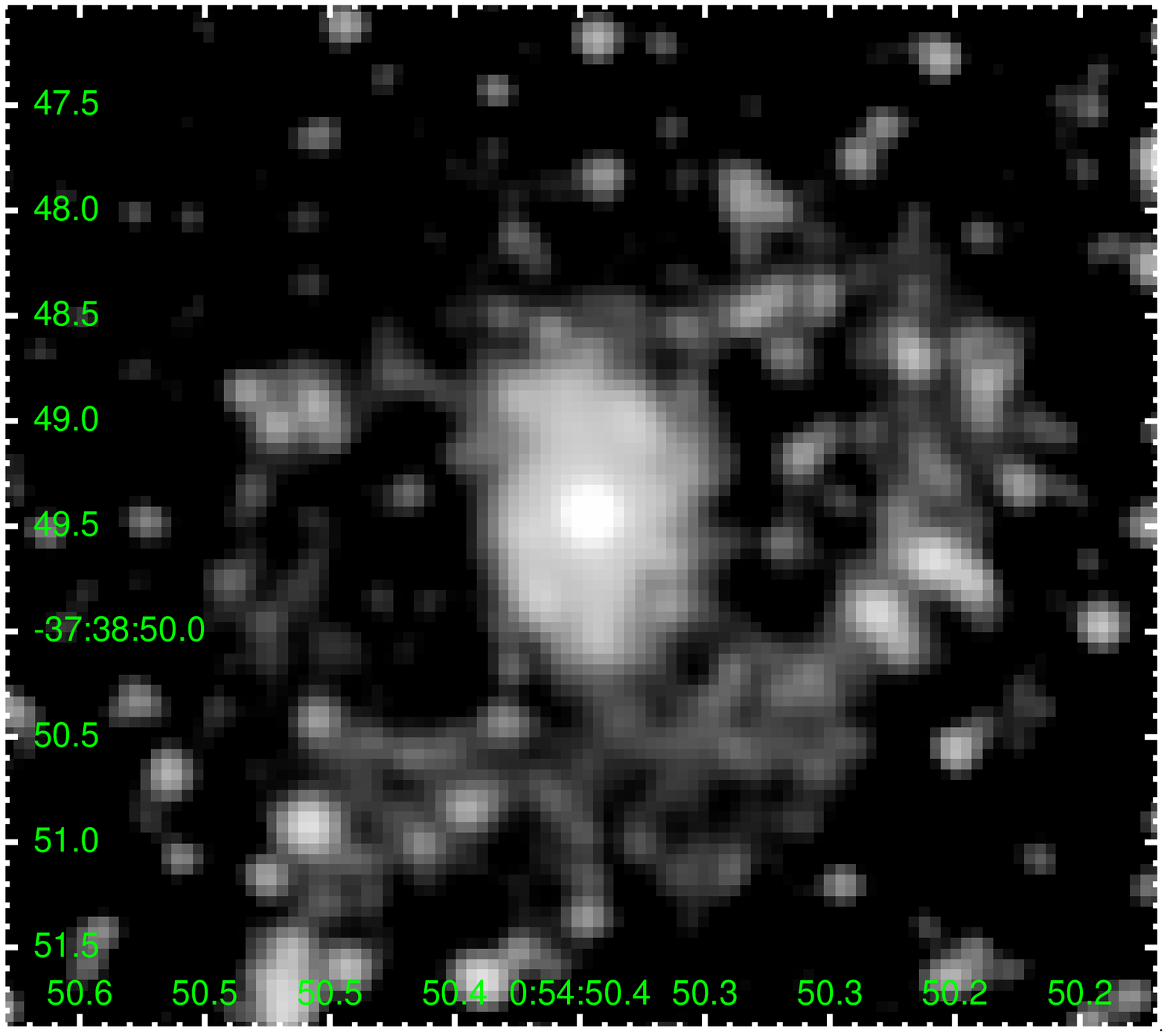}
\includegraphics[width=5.8cm,angle=0]{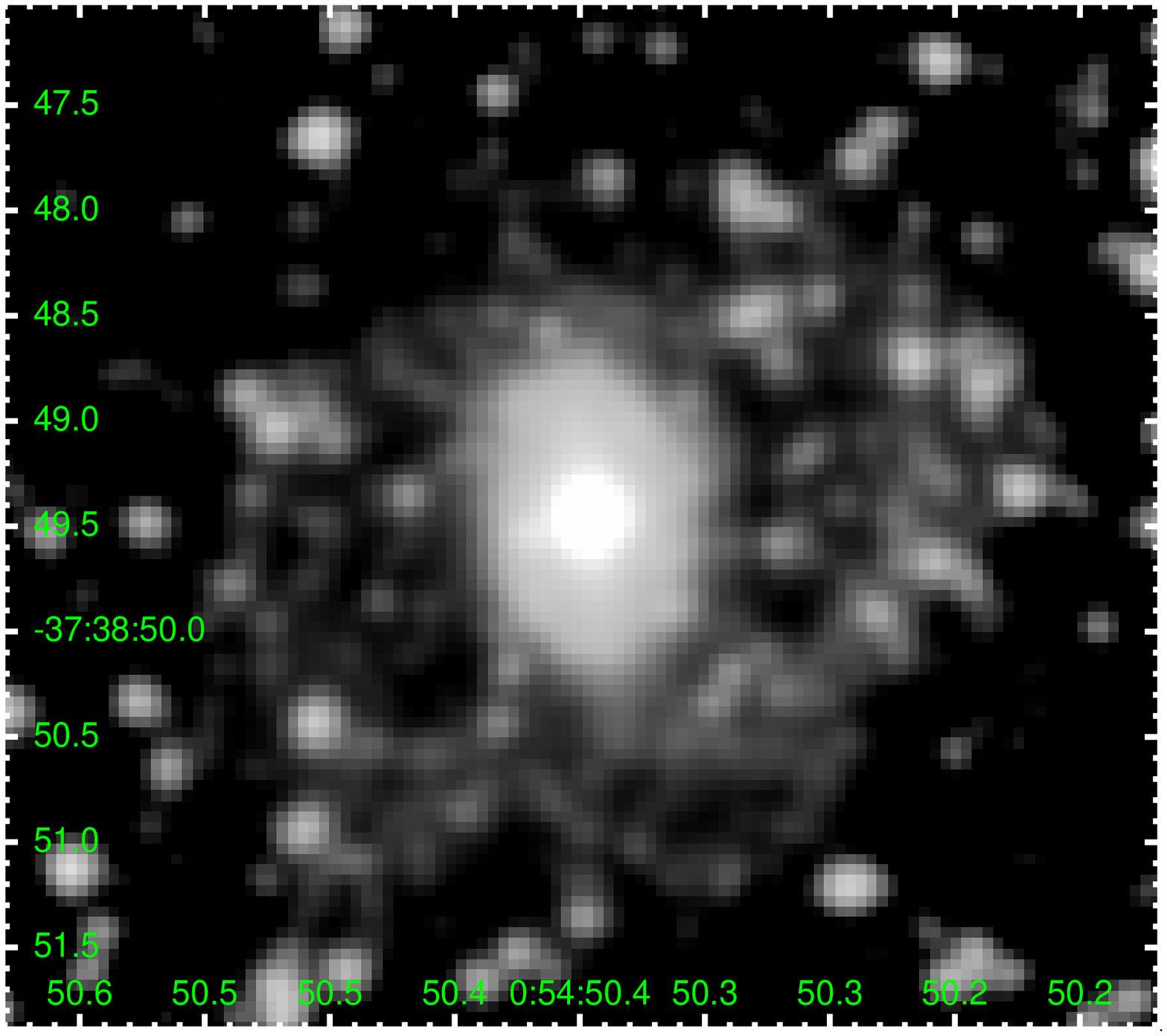}
\caption{HST images of the extended optical source positionally coincident with 3XMM J005450.3$-$373849. From left to right we show the F435W, F555W and F814W filters. X-ray contours are overlaied in white (left panel).}
\label{fig:imgHST}
\end{figure*}

\subsection {Optical Images and Photometry}

Figure~\ref{fig:imgHST} shows the HST images of the field surrounding 3XMM J005450.3$-$373849 in the aforementioned three filters. The Chandra X-ray contours are overlaied in Fig.\,\ref{fig:imgHST}, left panel. As it can be seen, the X-ray source is positionally coincident with the geometrical centre of the extended optical source, whose morphology clearly matches an R$'_1$-type ring galaxy, in which two outer pseudorings are conspicuous. These features are thought to arise from gaseous rings that developed at the outer Lindblad resonances in a spiral galaxy \citep[e.g.][]{BC1991}. The optical source also displays a barely resolved nuclear component (which we associate with a bulge plus an unresolved nucleus), located at the centre of an ``oval'', with hints of spiral structure, and two ``hot-spots'', clearly seen in the $B$-band image at position angles $66^\circ$ and $238^\circ$ (i.e., almost at diametrically opposite sides of the nucleus).

The inner disk (the ``oval'' feature) has an ellipticity $\epsilon = 0.36$
and a position angle $PA \approx 5^\circ$, spanning $0.86''$ and $0.57''$
along its major and minor axes, respectively.  The pseudorings, in
turn, present an angular size of $\sim 3.5'' \times 2.5''$ on the plane of
the sky.

\subsection {Optical spectroscopy}

\begin{figure}
  \centering
\includegraphics[width=0.85\columnwidth,angle=270]{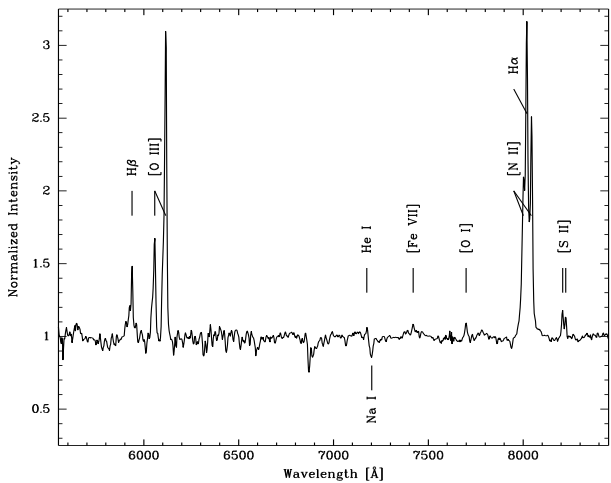}
  \caption{Gemini-GMOS spectrum of the nuclear region of the optical counterpart of 3XMM J005450.3--373849. Prominent emission and absorption lines are indicated.}
  \label{sp}
\end{figure}

In Fig.~\ref{sp}, we show the GMOS spectrum over the spectral range 5500--8500 \AA . Emission lines of H, \ion{He}{i}, [\ion{N}{ii}], [\ion{O}{i--ii--iii}], [\ion{S}{ii}], [\ion{Ar}{iii}], and [\ion{Fe}{vii}] were identified in the spectrum and measured for radial velocities. We obtained a mean $z=0.2216\pm0.0008$ (13 measurements). We have also identified some absorption lines of \ion{Ca}{ii}, \ion{Na}{i}, and the G-band. Five absorptions gave a $z=0.2223\pm0.0005$ (5 values averaged), thus they are originated at the same distance as the emission lines. 

The spectrum presents hydrogen and forbidden lines of the same width with a mean $FWHM\sim500\pm150$ km~s$^{-1}$, compatible with a Seyfert~2 galaxy spectrum. Also, the ratio H$\beta$ and [\ion{O}{iii}] $\lambda$5007 is 0.1 \citep[typical in Sy~2, following][]{weedman77}.

We also measured the $FWHM$ of the emission lines in the spatial dimension. All of them resulted similar to the $FWHM$ of the whole trace of the spectrum, thus the spectral source should be considered as point-like (at least, considering the seeing-limited Gemini's spatial resolution).

\section {Discussion}

Through this paper, for the first time, we present a deep study of the X-ray source 3XMM J005450.3$-$373849 an its optical counterpart to unveil its real nature. The information gathered in the previous sections from the optical and X-ray analysis shows that 3XMM J005450.3$-$373849 poses typical characteristics of a Seyfert-2 AGN galaxy, located at $z=0.222 \pm 0.001$ behind NGC\,300, as demonstrates the optical spectroscopic observations that we obtained with Gemini. The X-ray source is positionally coincident with the central region of an extended optical source detected with HST showing a complex structure, that clearly matches the morphology of a R$'_1$-type spiral galaxy with a conspicuous outer pseudo-rings structure. The positional correlation strongly suggests a physical association between both sources. We also note that this galaxy is unrelated to either of the two clusters detected in the background of NGC\,300 (Schirmer et al. 2003), with measured redshits $z=0.165$ (Cappi et al. 1998) and $z=0.117$ (Collins et al. 1995), respectively.

The object's redshift implies a luminosity distance $d_L= 1111 \pm 6$\,Mpc and a spatial scale 3.606 kpc/$''$ (we made use of Ned Wright's \emph{Cosmology Calculator}\footnote{\url{http://www.astro.ucla.edu/~wright/CosmoCalc.html}; \citealp{W-2006}}, with standard Cosmological parameters: $H_0=69.6$\,km\,s$^{-1}$\,Mpc$^{-1}$, $\Omega_\mathrm{M}=0.286$, and $\Omega_\Lambda=0.714$). The (projected) major axis of the pseudo-rings system is thus $\approx 12$\,kpc, while the surface brightness profile of the inner ``oval'' feature can be fit by an exponential with a scale-length of $0.56 \pm 0.02''$ ($2.02 \pm 0.06$ kpc) in the $I$-band (rest-frame $R$-band). The rings are resolved into several ``knots'' with apparent magnitudes ranging from $V \sim 25$ to $V \sim 27$ mag ($-15.2 \lesssim M_V \lesssim -13.2$) and relatively red colours ($0.0 \lesssim B-V \lesssim 1.5$). Thus, they are probably large star-forming complexes (the H$\alpha$ emission falls within the F814W passband at the object's redshift), significantly reddened by the intervening dust in the disc of NGC\,300.

Though a small fraction of observed rings may be due to collisions or mergers of galaxies, or to accretion of intergalactic gas, the vast majority of rings are probably simple resonance phenomena, caused by the actions of a rotating bar or other non-axisymmetric disturbance on the motions of gas clouds in the disc \citep[e.g.][]{Sch1981}. In particular, outer pseudo-rings are in fact spiral arms with a very low pitch angle, winding upon themselves following the outer Lindblad resonance (OLR). These features are quite common in Seyfert galaxies \citep{HM-1999}, and it has been suggested that galaxies with outer (pseudo)rings tend to display enhanced nuclear star formation \citep{BC1991}; however, the real dependence of AGN activity on the global structure of the host galaxy is still debated \citep {Cis+2015}.

The information obtained from the optical data are in accordance with the results of our temporal and spectral X-ray analysis of 3XMM J005450.3$-$373849, where we have shown that the source was in two different states, showing typical AGN variability. The X-ray spectra available were best-fitted using an absorbed power-law model, with absorption columns in the $0.33-0.75 \times 10^{22}$~cm$^{-2}$ range and spectral indices of 1.5--2 typical of AGNs. In the 2000--2001 period, the source showed the highest flux, reaching 0.12--0.14 $\times 10^{-12}$~erg~cm$^{-2}$~s$^{-1}$. Contrarily, during the 2005--2010 period, the source was fainter with a flux of $\sim$0.06 $\times 10^{-12}$~erg~cm$^{-2}$~s$^{-1}$ with a higher absorption, returning to a brighter state in the last 2010--2014 period. Taking into account the derived unabsorbed X-ray fluxes of the source, assuming a 1110~Mpc distance to the source, we compute X-ray luminosities in the $\sim$8$\times$10$^{42}$ to 2$\times$10$^{43}$~erg~s$^{-1}$ range for the 0.5--8.0~keV energy band, for any of the spectral models assumed, which correspond to a weak AGN with little obscuration  \citep{ueda2014,merloni2014}. It is interesting to note that considering an $E(B-V) = 0.075$ for NGC\,300 \citep{gieren2004} and assuming $R = 3.1$, we deduce an $A_V = 0.2325$. Thus, following \cite{foight2015} we obtain a total absorption column $N_{\rm H}= 5-7 \times 10^{20}$~cm$^{-2}$. Hence, the best fit values obtained for the absorption columns in both spectral models are at least a factor of 5 higher than the foreground material.

In any case, this particular galaxy has signatures of strong star formation activity, as well as of an AGN.  The two hot spots, as said, are prominent in the F435W image (which corresponds to the $U$-band at the source rest-frame). Such spots, roughly aligned on opposite sides of the nucleus, are associated to strong ongoing star-formation, as is the case of other well-studied starburst galaxies \citep[e.g., M\,94;][]{WFK-2001}.

\section*{Acknowledgements}

We thank the anonymous referee for her/his insightful comments and constructive suggestions that lead to an improved manuscript. JAC is a CONICET researcher. This work was supported by  Consejer\'{\i}a de Econom\'{\i}a, Innovaci\'on, Ciencia y Empleo of Junta de Andaluc\'{\i}a under excellence grant FQM-1343 and research group FQM-322, as well as FEDER funds. MJR was supported by grant PIP 112-201101-00301 (CONICET). RG was supported by grant PIP 112-201201-00298 (CONICET). FG and MJR are fellows of CONICET.




\bsp	
\label{lastpage}
\end{document}